# Autour du « Songe » de Kepler

Jean-Pierre Luminet, Laboratoire Univers et Théories
Observatoire de Paris-CNRS- Université Paris Diderot (France)
email : jean-pierre.luminet@obspm.fr

June, 15, 2011

**Résumé**

Johann Kepler (1571-1630) est parfois considéré comme un précurseur des romans de science-fiction avec l'écriture de *Somnium, sive opus posthumum de astronomia lunaris* [1]. Dans cet ouvrage publié à titre posthume en 1634 par son fils Ludwig, Kepler essaie de diffuser la doctrine copernicienne en détaillant la perception du monde pour un observateur situé sur la Lune. Il explique : « Le but de mon Songe est de donner un argument en faveur du mouvement de la Terre ou, plutôt, d'utiliser l'exemple de la Lune pour mettre fin aux objections formulées par l'humanité dans son ensemble, qui refuse de l'admettre. Je pensais que cette vieille ignorance était bien morte, et que les hommes intelligents l'avaient arrachée de leurs mémoires, mais elle vit toujours, et cette vieille dame survit dans nos Universités. »

Kepler n'est certes pas le premier à faire le récit fantastique d'un voyage de la Terre à la Lune pour faire « passer un message ». Mais son *Songe* se singularise sous de nombreux aspects. En premier lieu, son auteur figure parmi les plus grands génies de l'histoire des sciences. En second lieu, il constitue le « chaînon manquant » entre les textes d'imagination pure de Lucien de Samosate, au II$^e$ siècle, et les aventures appuyées sur les découvertes scientifiques d'un Jules Verne, à la fin du XIX$^e$ siècle. En troisième lieu, le texte complet présente une extraordinaire structure en récits emboîtés, construite au fil des ans à mesure que l'astronomie nouvelle progressait. Il faut aussi retenir le rôle dramatique que sa diffusion, bien que confidentielle du vivant de Kepler, a eu sur sa propre vie de famille, ainsi que l'influence très profonde qu'il a exercée sur tout un courant de la littérature spéculative axée sur le voyage spatial.

## Genèse de l'œuvre et influences

Le *Songe* a vraisemblablement été rédigé en 1609 – année particulièrement féconde dans la production keplerienne, qui a vu la publication de *Astronomia Nova,* où sont exposées les deux premières lois des mouvements planétaires, et *Strena sive de Nive sexangula*, où la forme sexangulaire des cristaux de neige et autres structures naturelles sont déduites de profondes considérations géométriques.

La source du *Songe* est la *Dissertatio* de 1593, qui porte comme titre *Comment les cieux apparaissent à un homme situé sur la Lune*. Kepler y admet par hypothèse que la Terre se meut rapidement sur elle-même, bien que ses habitants n'aient pas conscience de ce mouvement. Il raisonne alors par analogie et place un « observateur » sur la Lune. Celui-ci connaît une expérience identique à celle de l'homme sur la Terre : l'observateur lunaire, dans la mesure où il ne participe plus à la rotation terrestre, va voir la Terre changer de quartiers, tout comme les terriens voient les phases de la Lune.

La vaste culture de Kepler en humanités gréco-latines inscrit le récit du *Songe* dans la longue tradition littéraire des fictions issues de rêves (*Songe de Scipion* de Cicéron au I$^{er}$ siècle avant l'ère chrétienne, *Songe de Macrobe* au V$^e$ siècle, etc.). Le premier écrit connu abordant le voyage physique (et non pas mental) dans l'espace est *L'Histoire Vraie,* composée en 160 après J.-C. par Lucien de Samosate – que Kepler a par ailleurs traduit afin d'apprendre le grec. Lucien raconte comment la nef d'Ulysse, aspirée en mer par une effroyable tornade, a vogué sept jours à travers l'espace pour se poser enfin sur la Lune. Mais Lucien expédie son équipée céleste en quelques lignes, le véritable propos étant de faire une satire des historiens qui présentent comme véridiques des récits invraisemblables et mensongers. D'ailleurs, un autre récit de Lucien, l'*Icaroménippe*, raconte un voyage dans la Lune, mais là encore, à aucun moment le voyage relaté n'a recours à une technologie « vraisemblable », et Lucien ne le présente jamais comme réalisable à l'aide de la science.

Une autre fiction de l'époque entraînant le lecteur dans la Lune est le fameux *Du visage qui apparaît dans le rond de la lune* de Plutarque (vers l'an 100). L'auteur considère que la Lune est un astre identique à la Terre, susceptible d'être habité. Il suggère que la Lune possède des crevasses dans lesquelles le Soleil ne brille pas, et que les tâches sombres se trouvant sur le disque lunaire se sont que les ombres de rivières, de montagnes et de vallées.

Non seulement Kepler a traduit le traité de Plutarque, mais dans la première édition de son *Songe,* les deux textes se suivent. Kepler ne donne pas la raison de cette concomitance, mais elle incite à des hypothèses. La première renvoie à une époque de la pensée où les champs du savoir n'avaient pas la même ligne de partage que pour nous, et permettaient la coexistence d'un texte de spéculation merveilleuse et d'un autre de spéculation scientifique. La seconde laisse à croire que Kepler veut prudemment faire passer pour une fiction issue de rêves ce traité d'inspiration copernicienne, où de plus la « Surnature » est invoquée pour amener son observateur sur la Lune.

C'est à la même tradition du voyage imaginaire qu'appartient *La Divine Comédie* de Dante. Dans le Moyen-Age chrétien cependant, la traversée céleste a pour seul objectif de rejoindre le séjour divin, de sorte que le poète traverse le ciel sans le regarder... Les sphères planétaires que le Florentin croise en leur tournant le dos n'ont pour lui aucun intérêt ; comme dans une ascension mystique, la montée est d'une promptitude surnaturelle, le trajet d'un astre à l'autre presque instantané.

Le genre perdure en pleine Renaissance. Dans le chant XXXIV du *Roland Furieux* de Ludovico Arioste (1516), Astolphe, conduit par Saint Jean l'Evangéliste, emprunte le char d'Élie pour monter dans la Lune. Il y découvre un vallon dans lequel est rassemblé tout ce qui a été perdu sur terre. Un siècle plus tard, le chevalier Marino (*Adone*, 1623) enlève tout aussi facilement son Adonis dans le char de Vénus, et l'emporte dans les sphères célestes. Dans ces deux récits, la visite des mondes planétaires est bien plus intéressante que le vol lui-même - lequel se réduit à quelque chose d'aussi étriqué que le cosmos clos aristotélicien.

Cependant, la Renaissance donne aux esprits un nouvel élan ; sans que l'architecture du monde ait encore changé, l'attitude de l'Homme face au ciel se fait plus hardie, plus confiante. L'Homme a reçu en héritage le monde pour l'explorer et le dominer, dès cette vie. Un tel enthousiasme est proche de la tentation d'orgueil. Le philosophe Giordano Bruno y cède à la façon d'un explorateur intrépide (*De Immenso*, 1591). Il y exprime pour la première fois l'ivresse du vol, sans appréhension ni hésitation aucune, et la joie du voyage sans fin, sans espoir de retour.

Telles sont les influences littéraires de Kepler lorsqu'il compose la première version du *Songe*. Toutefois, à peine rédigé son récit, qu'il ne fait circuler que confidentiellement sous forme manuscrite, Kepler reçoit le choc du *Messager des Etoiles* (1610), dans lequel Galilée révèle ses découvertes télescopiques : le relief lunaire, les satellites de Jupiter, les amas d'étoiles, etc. Cette lecture ne fait que confirmer l'intuition de l'astronome allemand sur la nature terrestre des planètes. Il rédige aussitôt une lettre de soutien, publiée sous le titre de *Dissertatio cum Nuncio Sidero* (*Conversation avec le messager des étoiles*), puis après avoir lui-même observé ces satellites, il publie ses observations dans *Narratio de Observatis*

*Quatuor Jovis Satellibus* (c'est d'ailleurs Kepler qui, le premier, dans cet ouvrage de 1611, utilise le mot « satellite » pour désigner les quatre petits astres tournant autour de Jupiter).

Pour le visionnaire allemand, le vol interplanétaire est pour demain, et les pionniers ne manqueront pas : « Qui aurait cru autrefois que la traversée du Grand Océan était plus calme et moins dangereuse que la navigation dans les golfes étroits et traîtres de l'Adriatique et de la Baltique ? Créons des navires et des voiles adaptés à l'éther, et il y aura un grand nombre de gens pour n'avoir pas peur des déserts du vide. En attendant, nous préparerons, pour les hardis navigateurs du ciel, des cartes des corps célestes ; je le ferai pour la Lune et toi, Galilée, pour Jupiter », écrit-il avec enthousiasme.

C'est à partir du *Messager des étoiles* que Kepler va reprendre son *Songe*, le surcharger de notes, de gloses et d'un appendice, dont le volume excèdera largement celui du récit initial. Le texte complet, retravaillé jusqu'à sa mort en 1630, est donc un « work in progress » intégrant au fur et à mesure de son élaboration les découvertes du champ dans lequel il se situe.

### Le Récit

Le *Songe* est le récit d'un rêve que fait Kepler une nuit après avoir observé la Lune et les étoiles. Il dit avoir rêvé d'un livre qui parle d'une aventure vécue par un certain Duracotus.

Duracotus, par les hasards d'une vie baignée de magie et d'un peu de sorcellerie, se retrouve adopté par Tycho Brahé pour y apprendre l'astronomie. Devenu érudit, il retourne dans sa patrie pour y retrouver sa mère, Fiolxhilde qui, avant de mourir, lui révèle certains secrets, et plus particulièrement la possibilité qu'elle a d'invoquer certains esprits, notamment un démon qui permet une sorte de voyage astral particulièrement efficace. Avec lui, elle effectue un dernier voyage pour aller retrouver une île nommée Levania, qui n'est autre que la Lune. Seulement, pour s'y rendre, il faut remplir certaines conditions :

« A une distance que cinquante mille milles allemands dans les hauteurs de l'éther se trouve l'île de Levania. La route qui va d'ici à cette île ou de cette île à notre Terre est très rarement praticable. Quand elle l'est, il est aisé pour ceux de notre race de l'emprunter, mais il est extrêmement difficile de transporter des hommes et ils risquent leur vie.

Nous n'admettons personne qui soit sédentaire, ou corpulent, ou délicat; nous choisissons ceux qui passent leur vie à monter les chevaux de chasse ou vont fréquemment aux Indes en bateau, accoutumés à se nourrir de biscuit, d'ail et de poisson fumé. Mais surtout nous conviennent les petites vieilles desséchées, qui depuis l'enfance ont l'habitude de faire d'immenses trajets à califourchon sur des boucs nocturnes, des fourches, de vieux manteaux. Les Allemands ne conviennent pas du tout, mais nous ne refusons pas les corps secs des Espagnols. »

Vient ensuite une description du voyage, la nécessité d'endormir le voyageur humain, de le droguer, les problèmes d'absence d'air à respirer (personne à l'époque ne savait qu'entre la Terre et la Lune, il manquait ce gaz indispensable à la vie), la nécessité d'une impulsion violente au départ mais la possibilité de laisser flotter le corps pendant la majeure partie du voyage (inertie) en n'oubliant pas le risque d'écrasement à l'arrivée. Kepler en profite pour se moquer des croyances de l'époque qui lui paraissent excessives. En fait, ils arrivent sur Levania en profitant d'une éclipse totale de Lune, seule possibilité pour les sorcières et les démons de faire le voyage dans cette nuit qu'est le cône d'ombre de la Terre. Mais l'éclipse ne dure pas plus d'une heure, il faut faire vite :

« Malgré sa longueur, tout le trajet se fait au plus en quatre heures. Nous sommes toujours très affairés et nous sommes d'accord pour partir seulement quand l'éclipse de Lune a commencé à l'Est. Si la Lune retrouve tout son éclat quand nous sommes encore en chemin, notre voyage devient alors inutile. Ces occasions si soudaines font que nous n'avons que quelques hommes pour compagnons, seulement ceux qui ont pour nous le plus de considération. Tous ensemble nous nous jetons sur un homme de cette sorte, tous nous le poussons par en-dessous et l'élevons dans les airs. Le choc initial est très pénible pour lui, il souffre comme s'il était un projectile lancé par un canon et voyageait au-dessus des mers et des montagnes. Il faut donc l'endormir dès le départ à l'aide de narcotiques et d'opiats, et déployer ses membres pour que l'avant de son corps ne soit pas séparé de l'arrière, ni sa tête du reste du corps, et que la violence du choc se répartisse dans chacun de ses membres. De nouvelles difficultés se présentent alors: le froid itense, et l'impossibilité de respirer. Nous remédions au premier en utilisant un pouvoir inné en nous, à la seconde en passant des éponges humides sous ses narines. Quand la première partie du trajet est accomplie, le transport devient plus aisé. Nous laissons alors les corps flotter à l'air libre et retirons nos mains. Les corps se mettent en boule comme des araignées, et nous les transportons presque par notre seule volonté, si bien que la masse du corps se dirige d'elle-même vers l'endroit prévu. »

Une fois parvenu sur la Lune, Kepler décrit des paysages fantastiques, où les montagnes sont beaucoup plus hautes que celles de la Terre, où les plantes poussent à des vitesses vertigineuses pour disparaître le même jour, un monde où vivent des animaux ressemblant à d'énormes reptiles. Il les voit faire le tour de la Lune en d'immenses hordes, courant sur leurs longues pattes ou volant à l'aide de grandes ailes, ou même suivant les cours d'eau sur des bateaux. Leur peau spongieuse peut se dessécher en un seul jour et tomber en écailles...

Après la description des données astronomiques liées à la Lune, à ses divers mouvements et particularités, à sa géographie (qui est le véritable objet du livre puisque Kepler l'a sous-titré *Astronomie lunaire*), l'auteur imagine la tête des habitants de la Lune, toujours cachés sur la face non visible depuis la Terre, dans des cavernes qui les abritent de l'extrême froid de la nuit (qui dure deux semaines) et de la chaleur intense qui règne pendant les deux semaines de jour. Certains habitants sont des plongeurs qui trouvent au fond des lacs lunaires cet abri face aux changements importants de température. Mais Kepler voit là son songe se terminer sans avoir pu finir le livre apparu dans son rêve…

A première vue, *Le Songe* de Kepler a l'air d'un voyage mythique, avec son peuple de démons volants, ses herbes magiques, ses commentaires astrologiques et son symbolisme médiéval ; en réalité, toute la description du voyage et de l'environnement lunaire (explicité dans les nombreuses notes) montrent que Kepler devance son époque d'au moins un siècle. D'une certaine façon il anticipe les voyages astronautiques. Les navigateurs de l'espace sont sélectionnés et préparés comme les cosmonautes d'aujourd'hui par des démons ressemblant aux actuels « sorciers » de la NASA. Tout cela n'aurait qu'un intérêt de curiosité si la science de Kepler, tout entière présente à l'arrière-plan du récit, ne se subordonnait au rêve.

### Structure du *Songe*

Outre sa situation entre un texte de Plutarque et les considérables ajouts en forme de notes, *Le Songe* se caractérise par une fascinante structure par emboîtements (qui sera plus tard l'une des spécialités de l'écrivain argentin Jorge Luis Borges).

Le récit principal de 1609 n'occupe qu'une vingtaine de feuillets. Les 223 notes annexées – titrées *Notes sur le Songe, écrites les unes après les autres entre 1620 et 1630* – sont quatre fois plus volumineuses et donnent des détails biographiques et astronomiques, sans plus aucune fantaisie. Un *Appendice géographique*, sous-titré *ou, si l'on préfère, Sélénographique*, tient une page et demie, qui se multiplient par dix en tenant compte des 39 nouvelles notes, numérotées cette fois de A à Z et de Aa à Mm. La seule note A contient XXXIV commentaires scientifiques !

Le récit proprement dit se déploie sur trois niveaux narratifs, dont l'interaction est significative :
- un narrateur à la première personne, historiquement situé en 1608, lit des chroniques et s'endort dans un grenier. En rêve il lit un autre livre, qui provient de la foire de Francfort ;
- dans ce livre, un nommé Duracotus prend le relais. Il conte ses aventures d'adolescence, qui le mènent dans le champ du savoir astronomique. Il retourne chez sa mère, qui lui fait rencontrer un étranger (un démon), lequel entame un récit ;
- il s'agit du récit d'un voyage vers la Lune par des moyens magiques (impulsion des démons), mais qui tiennent compte pourtant de difficultés techniques. Ensuite, depuis la Lune, nous avons une description du système solaire dans l'optique de la *Dissertatio*, sans oublier enfin la description extraordinaire des habitants du lieu ;
- retour inopiné au niveau 1 par le réveil du dormeur sous la pluie : nous ne connaîtrons pas la fin du livre inséré.

Ensuite, le récit est complété de notes, nous l'avons vu beaucoup plus importantes en volume que le récit lui même. Ajoutées au fil des ans après que Galilée eut utilisé la lunette, ces notes dialoguent avec le savant italien, le lecteur ou avec Aristote, donnent des explications, dessinent de complexes diagrammes géométriques, formulent des hypothèses. Kepler y anticipe une connaissance précise des obstacles soulevés par un voyage vers la Lune, et même si la technologie du XVII$^e$ siècle ne peut y apporter une solution, il pense qu'il est théoriquement possible aux hommes d'atteindre la Lune.

Le texte mêle donc la fantaisie et la démonstration scientifique, le savoir neuf et les croyances populaires. Par exemple, la description des mouvements de la Terre dans le ciel est fondée sur l'hypothèse héliocentrique de Copernic, et appuyée sur des calculs qui construisent un modèle scientifique. L'étrangeté naît de la logique d'une modélisation neuve de la réalité, où l'on est obligé de croire son cerveau à défaut de croire ses yeux. Il en va de même dans la description des Luniens, où ce n'est plus sur des calculs que Kepler se fonde mais sur des analogies, ainsi que les notes le montrent. Si Kepler met des cavernes sur la Lune, c'est pour que les habitants puissent se protéger de la chaleur et du froid, qui alternent (3 notes explicatives : 214, 217, 220). S'il ajoute « tout ce qui pousse sur terre, ou vit sur terre est d'une taille monstrueuse. La croissance est très rapide ; rien ne vit longtemps puisque tout, êtres et plantes atteint une taille gigantesque », il explique : « il existe un rapport entre le mouvement lent des fixes pour nous et les rotations rapides de la Terre. Le même rapport me paraît exister entre la durée de la vie humaine et la taille réduite de nos corps. Par conséquent, sur la Lune, où le jour est 30 fois plus court que le nôtre, j'ai pensé qu'il fallait attribuer aux êtres vivants une vie brève et une croissance extrêmement rapide » (note 213). Il se refuse à faire de sa Lune une simple duplication de la Terre pour y bâtir une utopie à la façon de Thomas More, comme il se refuse les facilités d'une imagination délirante à la manière de Lucien. Sa Lune et ses habitants sont fondés sur des analogies à base d'hypothèses, comme le montre la note 209 : « Ce n'est pas une pure invention » ou la note 211 : « c'est un pur raisonnement. »

Une caractéristique du style narratif de Kepler – que l'on trouve d'ailleurs dans l'ensemble de ses écrits – est l'usage avoué de la plaisanterie : « Je plaisante les mœurs barbares des Ignorants » (note 10) ; « Sous la plaisanterie, la physique » (note 55), « Je me laisse ici aller à plaisanter » (note 56), « sous le voile de la plaisanterie, il y a aussi l'idée suivante » (note 61), etc.

L'alliance de l'allégorie et de la plaisanterie, caractéristique de la « pensée paradoxale », est spécifique du XVI$^e$ siècle : en témoignent au moins l'*Utopie* de Thomas More (1516) et l'*Éloge de la folie* d'Erasme (1511), admirés par Kepler.

**Le procès en sorcellerie**

Une redoutable vague de chasse aux sorcières sévit en Europe au début du XVIIe siècle, en particulier dans les régions protestantes. Les dénonciations sont fréquentes pour assouvir des haines personnelles, et les tribunaux sont prompts à user de la torture pour obtenir des aveux aussi détaillés que fallacieux. Les victimes des procès en sorcellerie sont à 80 % des femmes, appartenant en majorité aux classes populaires – par conséquent illettrées et incapables de se défendre. C'est ainsi qu'en 1615, tandis que Kepler travaille à son ouvrage *Harmonices mundi*, sa mère Katharina, alors âgée de 68 ans, est accusée de pratiques sataniques et de sorcellerie dans sa ville natale de Leonberg, dans le grand-duché de Wurtemberg. Katharina Kepler, née Guldenmann, que son fils qualifie lui-même de « petite, maigre, sinistre et querelleuse », avait été élevée par une tante qui avait déjà fini sur le bûcher pour sorcellerie. L'affaire est sérieuse ; les autorités religieuses commandent d'emprisonner et de juger toute personne soupçonnée d'avoir commerce avec le diable, et le prévôt de justice de Leonberg s'y applique avec zèle : cinq « sorcières » de la petite ville de Leonberg sont brûlées dans l'année.

Kepler, mathématicien impérial auprès de Rodolphe II à Prague, se rend à plusieurs reprises dans le Wurtemberg pour assurer la défense de sa mère, et fait appel à l'aide juridique de la faculté de Tübingen. Katharina passe 14 mois enfermée et subit la question, d'autant que le prévôt accuse le fils de pratiquer lui-même la sorcellerie : *Le Songe*, qui a circulé sous forme de copies manuscrites, ne fait-il pas état de pratiques sataniques ? Les descriptions de Duracotus et Fiolxhilde sont aisément assimilables à celles de Kepler lui-même et de sa mère. De fait, Kepler se sentira en partie responsable du procès intenté à sa mère, comme il l'explique dans la note 8 du *Songe* : « mon livre n'en a pas moins présagé un désastre familial quand il commença à être connu. […] Je veux dire que ces propos ont été recueillis par des esprits qui ne sont que noirceur et voient partout de la noirceur. » L'astronome va en conséquence passer six années à étudier le droit canon et rédiger plaidoirie sur plaidoirie devant les tribunaux.

Chose incroyable pour cette époque où l'issue de tels procès était systématiquement la mort et où la famille devait s'estimer heureuse de n'être pas elle-même compromise, les efforts de Kepler seront récompensés : Katharina Kepler est finalement libérée de toute charge de sorcellerie le 4 octobre 1621, au moins en partie pour des raisons techniques, son fils ayant pu faire valoir le « non-respect des procédures juridiques correctes dans l'utilisation de la torture. » [2]

**Conclusion**

*Le Songe* de Kepler est un texte spéculatif appartenant à la fois au domaine du littéraire (puisqu'il donne lieu à un récit) et de la controverse philosophique et scientifique. Au XVIe siècle, les « fables » anticipent la possibilité d'existence de ce qui deviendra la Science, et tentent de lui constituer un corps mythique par ces fictions, alors qu'elle n'est encore perçue que comme une simple possibilité. Le paradigme scientifique étant en train de se constituer en tant que tel, les fictions à la fois s'y réfèrent, l'illustrent et aident à le constituer.

*Le Songe* de Kepler a influencé un grand nombre d'œuvres littéraires. Dès 1611, le poète métaphysique anglais John Donne rédige un *Conclave Ignatii*, satire de l'ordre des Jésuites dans laquelle Ignace de Loyola est rejeté de l'Enfer par Lucifer et chargé de coloniser la Lune. Donne se réfère explicitement à Copernic, Tycho Brahé, Galilée et « Keppler ».

Peu après la publication posthume du texte complet du *Songe* et de ses notes en 1634 paraissent coup sur coup *Man in the Moon* de Francis Godwin (1638) [3] et *The Discovery of a World in the Moone* de John Wilkins (1638). Suivront *Insomnium Philosophicum* de Henry

More (1647) (qui influencera Isaac Newton dans ses conceptions mystico-philosophiques sur l'infini), *L'autre monde* de Cyrano de Bergerac (vers 1650), *The elephant in the moon* de Samuel Butler (vers 1650*), Paradise Lost* de John Milton (1667), *Iter Extaticum* du Père Athaniasius Kircher (1671) – qui a pour originalité de décrire un voyage dans le système du monde de Tycho Brahé –, et les *Entretiens sur la pluralité des mondes* de Fontenelle (1686).

Au XVIII[e] siècle, le vol cosmique, devenu thème littéraire à part entière, s'enrichit de mille parures. Dans l'intervalle il y a eu Newton, et l'espace newtonien, ouvert sur l'infini, est propice au voyage. Les *Nuits* d'Edward Young (1742) mettent au point les procédés classiques de l'itinéraire céleste ; il s'agit d'un voyage éducatif, à fin essentiellement religieuse. Au siècle suivant, le romantisme fait naturellement sien la notion d'infini. C'est à Jean-Paul Richter que l'on doit les chefs-d'œuvre poétiques du voyage cosmique (*Hesperus*, 1792, et *La Comète*, 1820). La fin du XIX[e] siècle est une nouvelle période de confiance ingénue en la science, qui voit le voyage cosmique remis à l'honneur, mais dans un univers assez différent où poudroient les brouillards mystérieux de lointaines nébuleuses. L'influence du *Songe* de Kepler se fait à nouveau sentir dans *Mondes imaginaires et mondes réels* de Camille Flammarion (1864), *Autour de la Lune* de Jules Verne (1869) ou *Les premiers hommes dans la Lune* de Herbert George Wells (1901), ces deux derniers ouvrages marquant les véritables débuts de la littérature d'anticipation, plus tard baptisée science-fiction.

**Références**

[1] Traduction française : *Le Songe ou Astronomie lunaire*, par Michèle Ducos (Nancy : Presses universitaires de Nancy, 1984)

[2] Cet extraordinaire épisode est développé dans Jean-Pierre Luminet, *L'œil de Galilée*, JC Lattès, 2009.

[3] Traduit et présenté par A. Amartin, Presses universitaires de Nancy, 1984.